# Intrinsic triple degeneracy point bounded by nodal surfaces in chiral photonic crystal

Dongyang Wang[1*], Hongwei Jia[1,2], Quanlong Yang[3], Jing Hu[1], Z. Q. Zhang[1], C. T. Chan[1*].

1. Department of Physics, Hong Kong University of Science and Technology, Hong Kong, China.

2. Institute for Advanced Study, Hong Kong University of Science and Technology, Hong Kong, China.

3. School of Physics and Electronics, Central South University, Changsha 410083, Hunan, China.

*Correspondence to: wangdy@ust.hk; phchan@ust.hk

**Abstract**

In periodic systems, band degeneracies are usually protected and classified by spatial symmetries. However, the $\Gamma$ point at zero-frequency of a photonic system is an intrinsic degeneracy due to the polarization degree of freedom of electromagnetic waves. We show here that in chiral photonic crystals, such an intrinsic degeneracy node carries $\pm 2$ chiral topological charge and the topological characters is the same as a spin-1 Weyl point manifested as a triple degeneracy of two linear propagating bands intersecting a flat band representing the electrostatic solution. Such an intrinsic triple degeneracy point (TDP) at $\Gamma$ is usually buried in bulk band projections and the topological charge at photonic zero-frequency has never been observed. Here, by imposing space-group screw symmetry to the chiral photonic crystal, the Brillouin zone boundary is transformed into an oppositely charged nodal surface enclosing the $\Gamma$ point. The emergent Fermi-arcs on sample surface are then forced to connect the bulk band projections of these topological singularities, revealing the embedded non-trivial topology.

Applying topological band theory[1, 2] to electromagnetic waves (EMWs)[3] allows for the exploration of both fundamental physics[4-10] and practical applications[4, 11, 12]. Topological photonic or phononic crystals have been used to demonstrate many interesting phenomena, such as negative refraction[13, 14], super-imaging[15], and chiral zero mode[16]. Potential applications have also been proposed and demonstrated such as topological lasers[17-27], which highlights the advantage of using topological ideas in optical devices.

The photonic Weyl degeneracy is of particular interest in the family of topological nodal points. They can be regarded as the EMW analogue of the unaddressed Weyl Fermion originally predicted as an elementary particle. These nodal points are momentum space magnetic monopoles and carry chiral topological charges and are responsible for the emergence of Fermi-arcs on sample surface that connect the projection of opposite charges[28]. In photonics, they appear as linear two-fold degeneracies in photonic bands and can be realized through breaking either the space-inversion (*P*)[7, 8, 29-34] or time-reversal (*T*) symmetry[35, 36] in a photonic crystal. The band crossings can be mathematically described with a 2×2 Hamiltonian using Pauli matrices as $H = \sum_i k \cdot \sigma_i$. Such a mathematical representation explains the stability of Weyl points, as changing system parameters only moves the Weyl point to a different position in momentum space, and their annihilation can only be in pairs of opposite charges.

Recently, the Weyl or Dirac points have been extended to more general forms that are manifested as higher dimensional band crossings[37]. For instance, the spin-1 Weyl point is a three-fold degeneracy that can be represented by the Hamiltonian $H = \sum_i k \cdot S_i$, where $S_i$ are the spin-1 generators. These unconventional nodal points have been explored in various platforms[38-52], and as multiband degeneracies, their appearance largely relies on the point or space group symmetries[37, 53-55]. However, an intriguing exception is the photonic Γ point at zero-frequency that is intrinsically degenerate[56, 57], where two fundamental transverse modes carrying orthogonal polarizations intersect. No symmetry is required to sustain this degeneracy. Furthermore, due to the existence of a zero-frequency electrostatic mode allowed in Maxwell's equations, the photonic Γ point becomes triply degenerate and thus provides a natural platform for generating spin-1 Weyl point[58, 59].

For metamaterials and photonic crystals with *PT* or $C_2T$ symmetries, the intrinsic photonic degeneracy[56, 57] is closely related to topological features such as the in-plane nodal chain or non-Abelian frame charges[60-62]. We highlight here that the intrinsic TDP can be transformed into an effective spin-1 Weyl point by simply introducing chirality to a photonic crystal[29, 63]. The full solutions of Maxwell's equations for a chiral media near the $k=0$ point exhibit six bands (see Supplementary Information 1), including two positive-frequency modes, two negative ones (unphysical), and two zero-frequency longitudinal solutions. Such a band dispersion can be understood as the overlapping of two spin-1 Weyl points that each incorporates one positive, one negative, and one longitudinal band. The spin-1 topological character of photonic bands can be understood through the helical basis of EMWs[58, 59], which however may only be observed when the modes degeneracy is lifted.

Here, by introducing chirality in photonic crystals to break circular polarization degeneracy, we show that the induced topological charge is embedded in the intrinsic TDP at Γ that behaves like a spin-1 Weyl point. Such intrinsic TDP is different from the previously observed finite-frequency Weyl points in chiral[33, 34] or chiral-hyperbolic materials[29, 30, 32] that are two-fold degenerate. It is also different from the spin-1 Weyl point that has been demonstrated[50, 51], which requires meticulous design to achieve the three-fold degeneracy of two linear bands and one flat band at a finite frequency. However, the topological charge at zero-frequency is usually obscured since the photonic Γ point is always buried in bulk band projections, and the opposite charge may appear at a random position in momentum space and the band structure projected onto a 2D plane may not have a band gap to sustain surface state arcs. Here, by further imposing screw symmetry in the space group, the two lowest propagating bands of the chiral photonic crystal will be symmetry enforced to be degenerate on the Brillouin zone (BZ) boundaries, appearing deterministically as an oppositely charged nodal surface[64-70] pinned at the zone boundary position that is far away from the zone centre. When such a Weyl point and nodal surface project onto the surface BZ together, the photonic surface state arcs will emerge to connect them. By direct experimental observation of such arcs, we reveal the embedded "+2" topological charge at the photonic Γ point of zero-frequency. Furthermore, a combination of screw symmetry and rotation symmetries can give rise to hourglass-like photonic band dispersions[56, 71], allowing for the existence of multiple Weyl points and nodal surfaces in higher bands that bring various types of surface state arcs in the surface BZ. Such topological arcs are then experimentally observed and their robustness in propagation is tested.

We constructed photonic crystals that exemplify such properties. We first consider a chiral metallic inclusion as shown in Fig. 1(a), where two helical resonators are related by a $C_{2z}$ symmetry. Such resonators induce a chiral response along the z-direction, coupling the electric component $E_z$ to magnetic component $H_z$ for an incoming EMW. The chiral response splits the equi-frequency surfaces of the two fundamental photonic modes at finite frequencies, leaving only the point node degeneracy at zero-frequency Γ point. Taking the electrostatic mode into account, the point can be recognized as a three-fold degeneracy of 2 linear bands and one flat band, and carries ±2 topological charge. It is different from the more common two-fold degenerate charge-2 Weyl points that have quadratic dispersion (e.g., the one at 8 GHz at Γ in the experiment shown below).

To directly observe the photonic surface state arcs induced by such intrinsic TDP, we construct a band degeneracy surface at BZ boundaries to carry topological charge as well. A screw symmetry $S_{2x}$ (or $S_{2y}$) is implemented to the resonators in Fig. 1(a), and a second copy is obtained as shown in Fig. 1(b). The blue resonators shift from the red ones by a half unit of [$p_x/2$, $p_y/2$, $p_z/2$]. The displacement along z-direction can be seen from Fig. 1(c). By putting such resonators into a periodic arrangement of $p_x = p_y = p_z = 6$ mm, we obtain the final unit cell as in Fig. 1(d). The photonic crystal has the space group P4$_2$2$_1$2 (No.94), where the screw symmetries of $S_{2x}$ and $S_{2y}$ confines the bands as two-fold degenerate at the BZ boundaries of $k_x = \pi/p_x$ and $k_y = \pi/p_y$ planes in Fig. 1(e).

The band structures of the proposed chiral photonic crystal are numerically calculated (CST Microwave studio) and shown in Fig. 1(f), with the band crossings marked in color to indicate the position and charge of Weyl nodes. The Intrinsic TDPs (purple) is located at the crossing between lowest two bands at zero-frequency and Γ point. Additional spin-1/2 Weyl point (two-fold degeneracies) can be found at Γ point for the 3$^{rd}$ and 4$^{th}$ bands, which is due to the $C_{2z}$ symmetry protection. Line degeneracies can be found along X-M-Y and Z-Z' for 1$^{st}$/2$^{nd}$ and 3$^{rd}$/4$^{th}$ bands that are protected by the screw symmetry and indicate the formation of nodal surfaces in momentum space. Moreover, an hourglass shaped band dispersion can be found in Fig. 1(f) for 1$^{st}$/2$^{nd}$ to 3$^{rd}$/4$^{th}$ bands, where the $C_{2z}$ symmetry is responsible for the degeneracies at Γ, and the screw symmetry protects the degeneracies at BZ boundaries. Such an shape of band dispersions leads to the crossing between 2$^{nd}$ and 3$^{rd}$ bands[56], which are charge-1 Weyl points (red and blue). The wave vector coordinates of these Weyl points and nodal surfaces are retrieved and presented in momentum space of Fig. 1(e), where four "+1" charge Weyl point are shown in red and four "-1" charge Weyl points are shown in blue. Charge "+2" TDP and quadratic Weyl point at the Γ point are marked in purple, and the charge "-2" nodal surface is shaded grey. These topological charges can be numerically retrieved by examining the Berry phase windings around the nodal structures (Supplementary Information 2). The topological structures are robust against geometrical parameter tunings in the chiral photonic crystal, provided that the chiral response and symmetries remains intact (Supplementary Information 3).

To explore the topological surface modes induced by the topological singularities, we examine the projected band structures. The sample configuration is shown in Fig. 2(a), where the sample is truncated along the z-direction and formed by stacking plates (periodic in the x-y plane) along the vertical direction. We calculated the band projections of the structure as shown in Fig. 2(b) and marked the surface mode bands in red. These surface modes can be classified into three frequency regions of A, B and C that span three projected band gaps from low to high frequency. For each of these frequency regions, we show the associated topological singularities in the surface BZs of Fig. 2(c). At the frequency range A, the surface modes appear inside the first band projection gap (between the projection of $1^{st}$ and $2^{nd}$ bands), where the topological charge carried by the intrinsic TDP at $\Gamma$ is "+2" and the charge is "-2" for the nodal surface at BZ boundaries. For the middle frequency range B, the surface modes appear in the second band projection gap ($2^{nd}$ and $3^{rd}$). The topological charges are carried by four pair of opposite Weyl points that project on the surface BZ as shown in the middle panel of Fig. 2(c). As to the high frequency region C, the surface modes are found in the third band gap ($3^{rd}$ and $4^{th}$). Topological charges in this gap are again carried by the charge "+2" (quadratic) Weyl point at $\Gamma$ point (finite frequency) and charge "-2" nodal surface at BZ boundaries.

These surface modes from different frequency regions can be experimentally observed through field mappings on the sample surface. The chiral photonic crystal is fabricated using Printed Circuit Boards (PCBs) techniques and a microwave measurement is conducted to characterize the sample and retrieve the projected bands, as schematically shown in Fig. 2(a). The measured results are shown in Fig. 2(d), where the predicted surface modes are observed in all three frequency regions, and we found good agreement with the theory prediction shown in Fig. 2(b).

We then study the topological surface state arcs on the sample surface. The arcs become evident as we show the computed and measured photonic equi-frequency contours (EFCs), which reveal their chiral character and relationship to the topological charges. The numerically calculated and experimentally measured results are shown in Fig. 3 for each of the above discussed frequency regions. In Fig. 3(a), the EFCs at frequency region A are shown, where two surface state arcs are found to connect the bulk band projections between surface BZ centre and boundary. The number of arcs reflects the topological charges carried by the TDP at $\Gamma$, as well as the nodal surface at BZ boundary, which thus verified the topological character of intrinsic TDP at photonic Γ point of zero-frequency. In Fig. 3(b), the EFCs at frequency region B is shown. Since there are four pair of charge-1 Weyl points, four Fermi-arcs are found to connect the projection of them as expected. In Fig. 3(c), two Fermi-arcs are again found to spawn from the surface BZ centre to the boundary, such Fermi-arcs connect the projections of quadratic charge-2 Weyl node at Γ and nodal surface at BZ boundaries that are embedded in the 3$^{rd}$ and 4$^{th}$ bands.

Chiral topological modes should be able to support robust propagations. We proceed to experimentally explore the propagation behaviour of the surface state arcs presented in Fig. 3. In Fig. 4(a-c), we show the measured field patten on the x - y plane of the sample surface (483 × 483 mm$^2$) which reveals the topological surface modes in three frequency regions. The source antenna is put at the centre of sample bottom edge in the x - y plane (x ≈ 240 mm, y = 0 mm), so that the surface modes in half-surface BZ ($k_y > 0$) are excited. The excited surface beams in various frequency ranges can be seen visually in Fig. 4(a-c), as a result of coupling the microwave wave into topological surface arcs. The EFCs shown in the insets explain the propagation of surface beams by indicating the direction of group velocities with arrows, which are normal to the tangent of the arcs. In particular, two beams $B_1$ and $B_2$ are excited in Fig. 4(b), which is due to the existence of multiple surface state arcs in the surface BZ and two of them (in upper half of surface BZ) are excited.

To examine the robustness of such topological surface modes, we further constructed the fabricated sample as a step structure with the step height of $h = 12$ mm (sample photo is shown in supplementary information 4). The source antenna is placed at the edge of higher surface, and we see whether the excited topological surface modes can propagate to lower surface regardless of the disruption by the step. In Fig. 4(d-e), the surface modes in first band gap (frequency region A) are shown for frequencies of $f = 6.1$ GHz and $f = 6.2$ GHz. We see that these surface modes can propagate from the higher surface to the lower one, uninterrupted by the step. We note that the step preserves $k_y$, and from the inset of Fig. 4(a), we see that the other propagation channel is at negative $k_y$, and hence no reflection can occur despite of the abrupt change in the step. For the second band gap (frequency region B) and at $f = 6.5$ GHz, the two surface beams $B_1$ and $B_2$ shown in Fig. 4(b) can be both excited. The propagations of these two beams across the step structure are shown in Fig. 4(f) and (g), respectively, where we see that their propagations still appear to be robust against the disruption. We note that a $k_y$-preserving propagation channel (with opposite group velocity) does exist for $B_1$, as can be seen in Fig. 4(b), but the reflection is still very weak for the same reason as surface modes in valley-Hall topological insulators are "topologically protected" (the reflection channel is far away in momentum space). In fact, the theoretical minor reflection for surface beam $B_1$ is hardly noticeable experimentally. In Fig. 4(h-i), the surface modes within third band gap (frequency region C) are verified for $f = 7.6$ GHz and $f = 7.2$ GHz. These surface waves can again propagate through the step and keep propagating on the lower surface due to a lack of reflection channel. Such results further demonstrate the transport behaviour of the surface waves on the chiral photonic crystal.

In conclusion, we have designed a chiral photonic crystal that demonstrates the topological character of intrinsic TDP at the zero-frequency photonic Γ point. The Γ point of the chiral crystal carries topological charge 2 (even though the dispersion is linear) and the crystal is also designed to carries topologically charged nodal surfaces at the BZ boundary that encloses the TDP. As the nodal surface is far away from zone centre, these degeneracy features allow for the direct observation of surface state arcs connecting the projection of these topological singularities revealing the embedded topological charges. This surface arc feature highlights the spin-1 Weyl point character of the zero-frequency Γ point when the crystal is chiral, as it is an intrinsic degeneracy point that does not require space group symmetry to sustain. Various topological surface state arcs are experimentally observed to support robust propagations on the surface of such chiral photonic crystals, which enrich the feasible schemes for realizing topological transport.

**Acknowledgments**

This work is supported by Research Grants Council of Hong Kong, China, AoE/P-502/20, 16310420, 16307621, and by the Croucher Foundation (CAS20SC01).

**Figures**

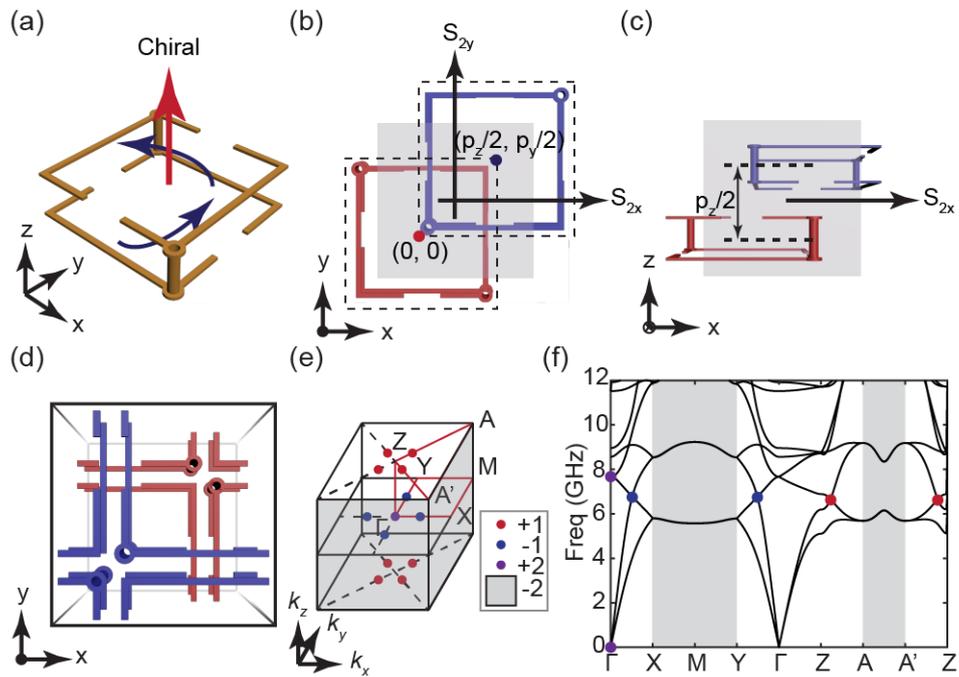

**Fig. 1. Chiral photonic crystal carrying degeneracy points and charged nodal surfaces.** (a) The metallic inclusions leading to chiral response along z-direction. (b) Two copies of resonators generated by screw rotations of $S_{2x}$ or $S_{2y}$. (c) Side view of the two copies of resonators. (d) The unit cell of the photonic crystal formed by periodic arrangement of resonators. (e) Weyl points and nodal surface in BZ. (f) Calculated band structures revealing the band degeneracies.

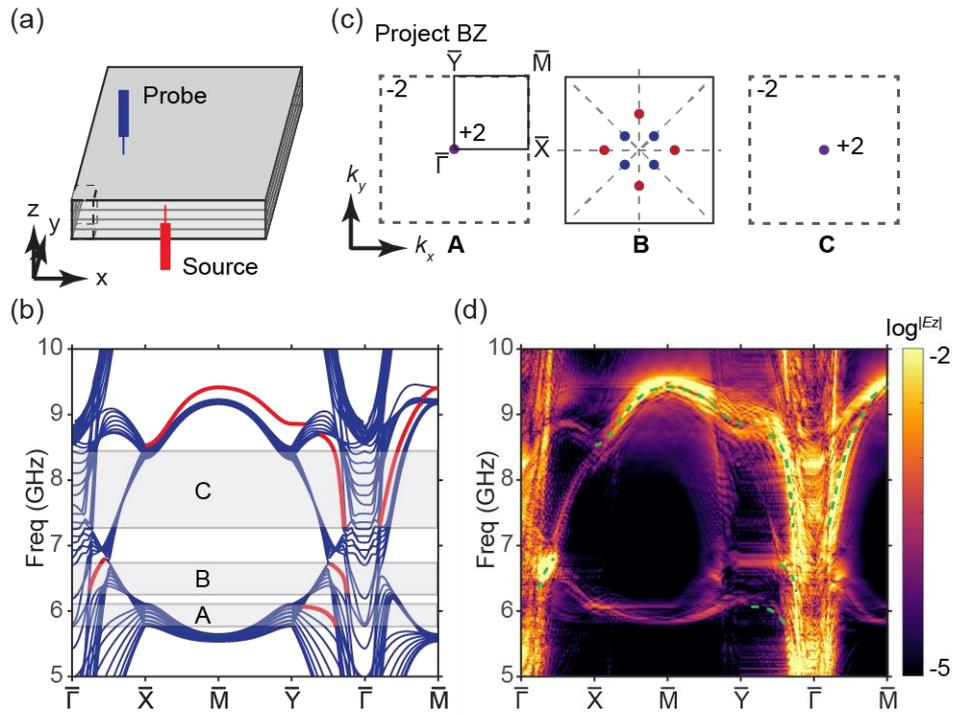

**Fig. 2. Topological charges and surface modes in band projection gaps.** (a) Schematic of the measured sample. (b) Simulation results of the projected band dispersions, surface modes are marked in red. (c) Surface BZ on x-y plane and the projection of topological charge for different bands, i.e., 1st and 2nd bands in A, 2nd and 3rd bands in B, and 3rd and 4th bands in C. (d) Experimentally measured band projections, and surface modes positions are indicated as dashed green lines.

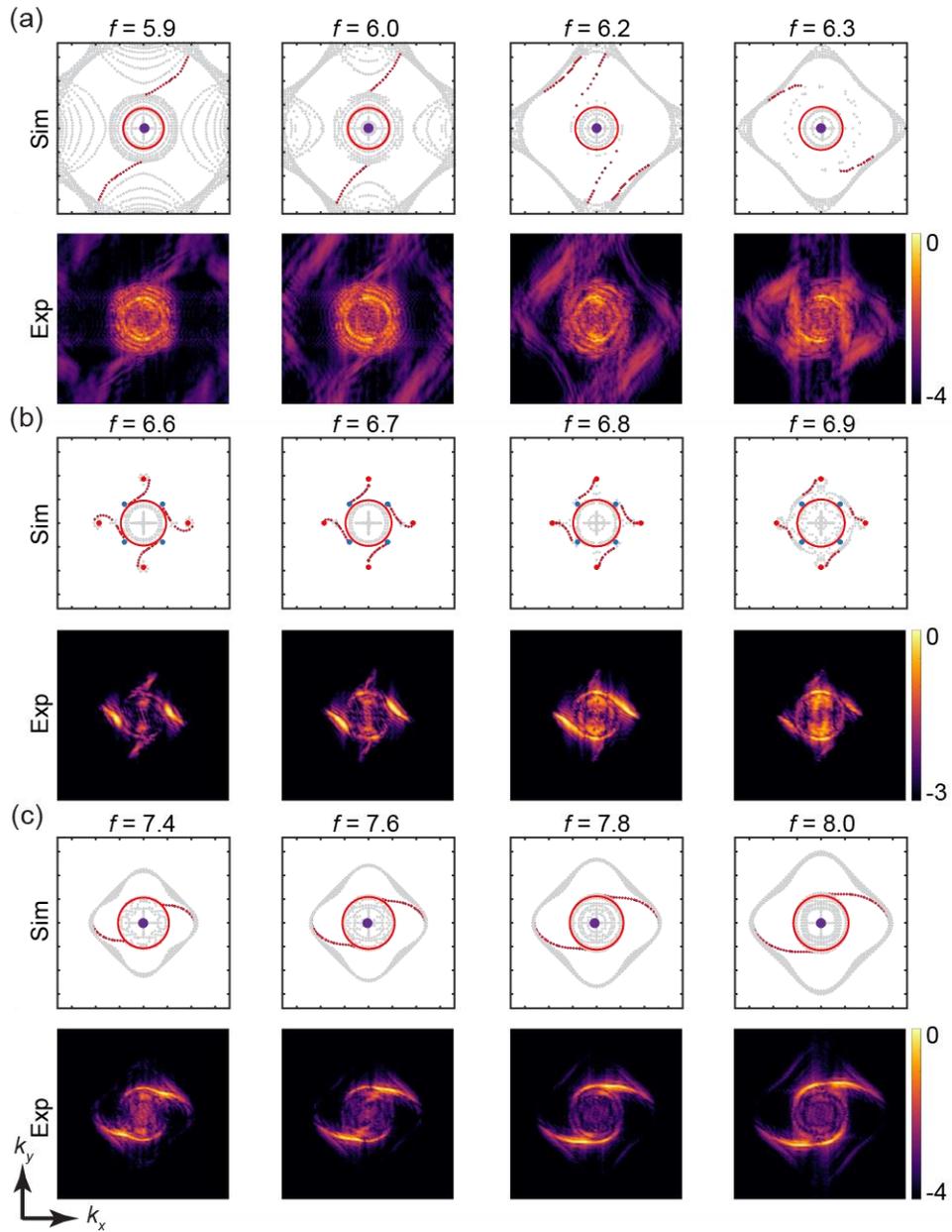

**Fig. 3. Chiral surface state arcs on the photonic crystal surface.** (a-c) The calculated and measured Fermi arcs for frequency ranges A, B, and C, respectively. The permittivity of substrate material in simulation is set as $\varepsilon = 1.7$ in (a), and $\varepsilon = 1.8$ in (b-c) to best fit the experimental results.

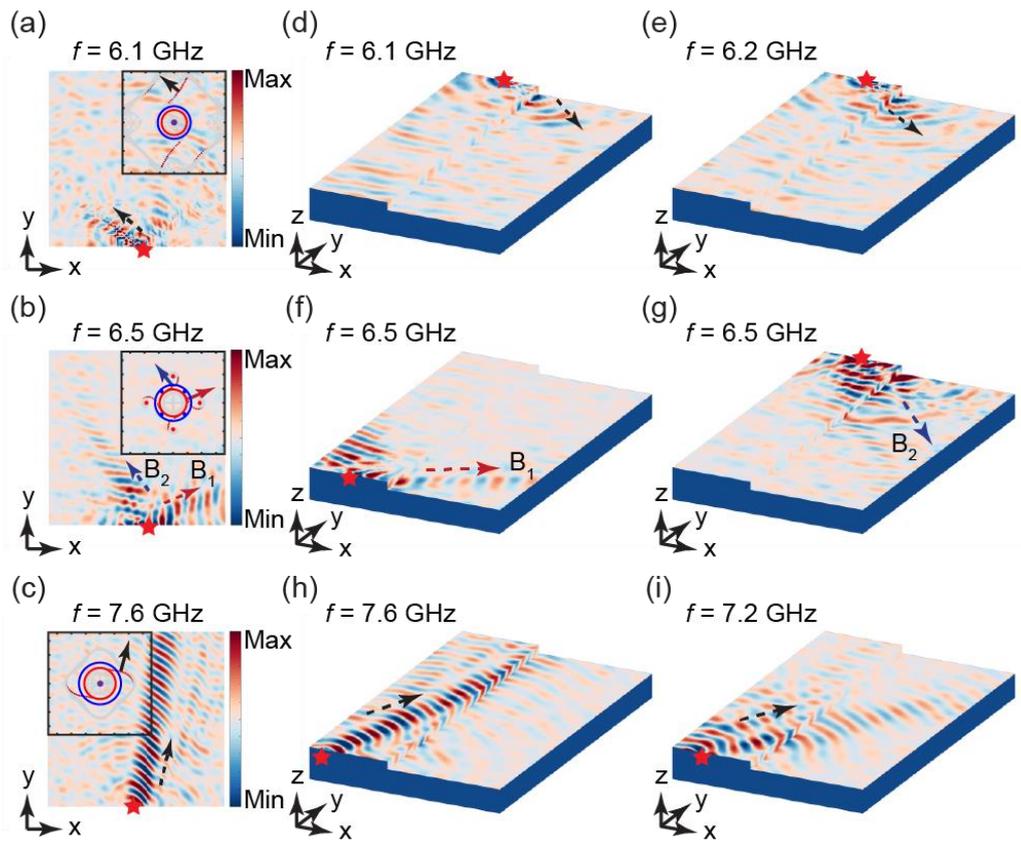

**Fig. 4. Reflectionless propagation from topological surface state arcs.** (a-c) The measured surface waves at frequency $f$ = 6.1, 6.5 and 7.6 GHz, respectively. EFCs are shown as insets. Source positions are indicated with red stars. No reflections can be found from the sample edges. (d-e) The propagation of EMWs at $f$ = 6.1 and 6.2 GHz on the step configuration of sample. (f-g) The propagation of EMWs at $f$ = 6.5 GHz for surface beams $B_1$ and $B_2$. (h-i) EMWs propagation at frequencies of $f$ = 7.6 and 7.2 GHz, respectively.